\documentclass[11pt,twoside]{article}

\usepackage{asp2006}
\usepackage{epsf}
\usepackage{psfig}
\usepackage{lscape}

\begin{document}
\title{Star clusters as diaries of galaxies}  
\author{Th. Maschberger and P. Kroupa}  
\affil{Argelander-Institut f{\"u}r Astronomie, Auf dem H{\"u}gel 71, D-53121 Bonn, Germany; Rhine Stellar Dynamics Network RSDN}   
\begin{abstract}
Most if not all stars form in star clusters.
Thus the distribution of star clusters preserves the information on the star formation history (SFH) of a galaxy.
Massive clusters form only during episodes of high star formation activity whereas periods of low star formation activity cannot produce them.
We present here the method of \cite{maschberger+kroupa} to derive the star formation history of a galaxy from its star-cluster content.
\end{abstract}

\cite{weidner-etal2004} showed that the mass of the brightest young (=most-massive) cluster in a galaxy follows from the present star formation rate (SFR).
The proposed scenario is that a complete (in a statistical sense) population of star clusters emerges during a {\it ``formation epoch''} which lasts for $\delta t \approx$ 10 Myr.
The total mass of the population is given by the star formation rate, $M_\mathrm{tot} = \mathrm{SFR} \times \delta t $.

A sample of most-massive clusters is characterised by the average mass of the most massive cluster, $\bar{M}_\mathrm{max}$, and is also parametrised by $M_\mathrm{tot}$, written symbolically $\bar{M}_\mathrm{max} = f(M_\mathrm{tot}) = f(\mathrm{SFR} \times \delta t)$.

The length of the formation epoch is determined from the observational data:
The brightest young cluster in a galaxy is normally the most massive one of the current formation epoch.
Independently the present star formation rate can be determined, e.g. from FIR or H$\alpha$ data.
With this a region which contains a certain fraction of the most massive clusters can be defined.
A fit of this region to the data then allows us to determine the duration of the formation epoch, since the height of the region depends on $\delta t$.
This gives $\delta t = 10$ Myr.

\begin{figure}[!ht]
\plottwo{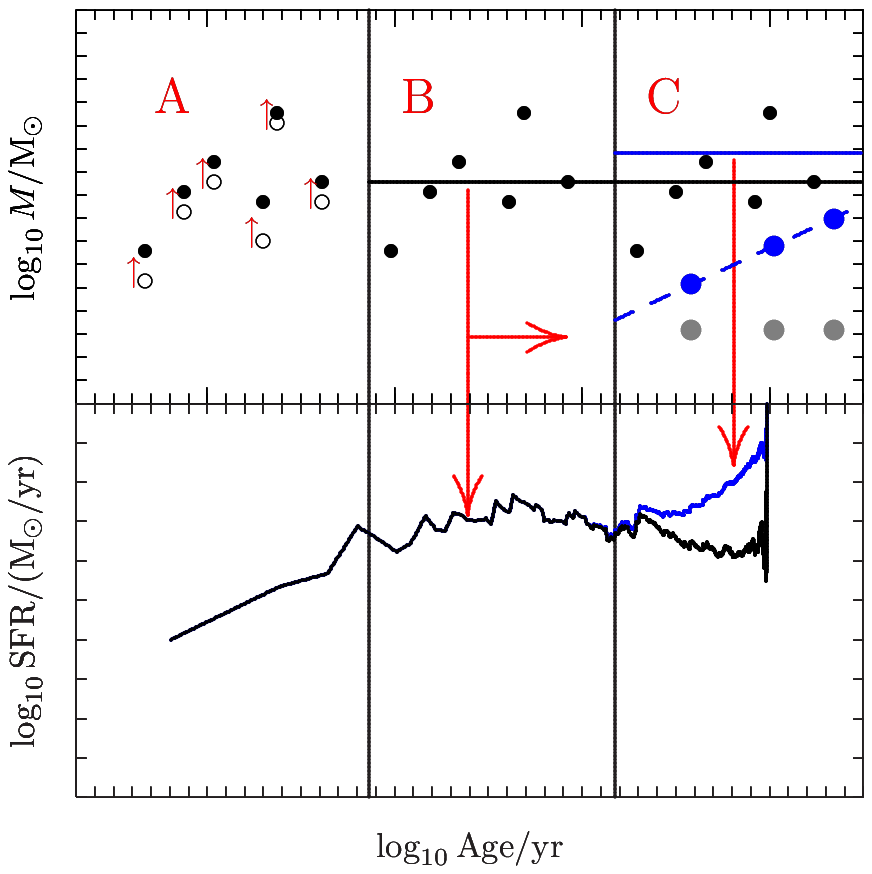}{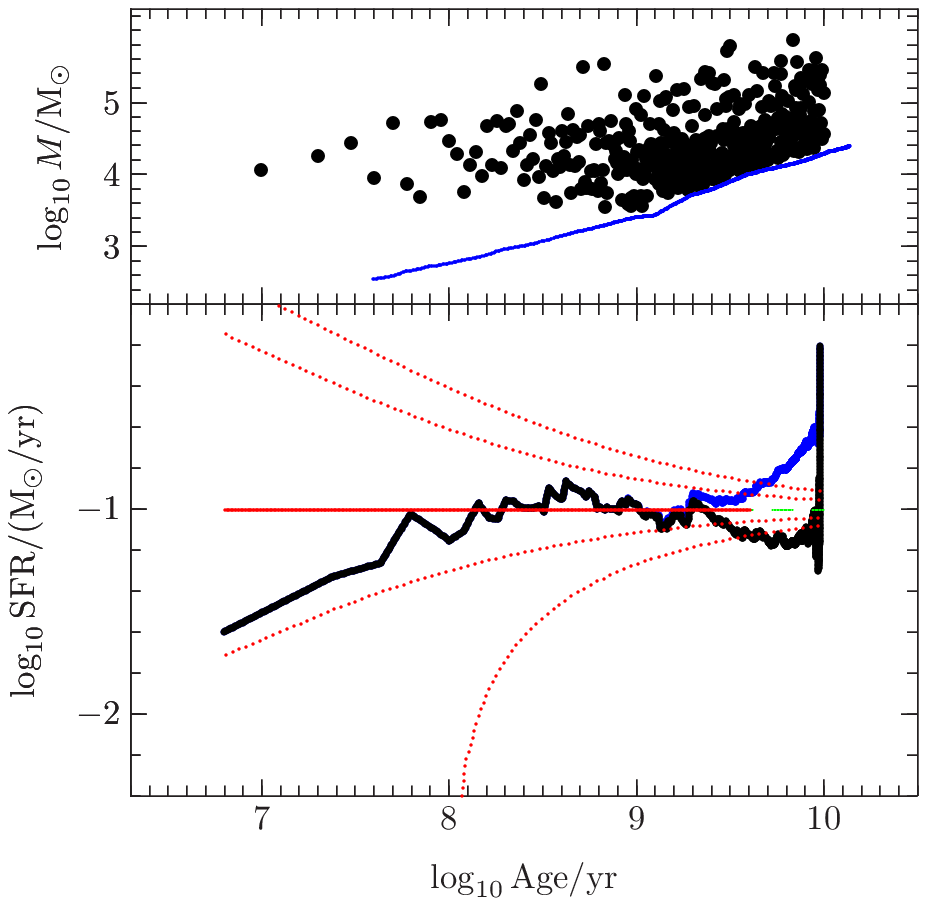}
\caption{Left: The method schematically. Right: Model for the LMC.}
\end{figure}

In this notion the SFH of a galaxy can be seen as a sequence of formation epochs, each containing a most-massive cluster.
This leads to the general concept to infer a SFH using the available relation between $\bar{M}_\mathrm{max}$ and the SFR (Figure 1, left panel):\\
\fbox{A}\quad 
Correct the observed cluster masses for dynamical evolution using e.g. the formulae of \citet{boutloukos+lamers2003}.\\
\fbox{B}\quad 
Calculate the average mass, $\bar{M}_\mathrm{max}$, of the most massive clusters in the averaging window.
Then use $ \mathrm{SFR} = f^{-1} ( \bar{M}_\mathrm{max} )$ to calculate the SFR.
Move the averaging window in time to obtain a continuous SFH.\\
\fbox{C}\quad 
Due to the evolution of star clusters and the observational limiting magnitude not every formation epoch contains a cluster. During such an epoch the SFR cannot be determined, but an upper and lower limit for it can be derived by assuming either no star formation ($M_\mathrm{max}=0$), giving the lower limit, or star formation is not detectable ($M_\mathrm{max}$ corresponding to the observational limiting magnitude), giving the upper limit.

To study the capabilities of our method we generated a model with parameters similar to those for the Large Magellanic Cloud: constant SFR of 0.1 M$_\odot$/yr (the total mass of the LMC is $2\times10^9$ M$_\odot$, \citealt{kim-etal1998}) 
and  cluster disruption parameter $t_4=7.9 \times 10^9$ yr, \citet{boutloukos+lamers2003}.

In the upper part of Figure 1,  right panel, the most-massive clusters are shown, together with the observational limiting mass (solid line, limiting magnitude -3.5 mag, \citealt{hunter-etal2003}).
The lower part shows the reconstructed SFH (thick solid lines), which branches in two parts at early times due to the observational limit.
The horizontal line shows a constant SFR, resulting from averaging up to the time where the line stops. 
If the reconstructed SFH would cross the red dotted lines, then we would conclude that the underlying SFH has variations.
But since this is not the case the reconstructed shape of the SFH agrees with the assumed constant SFR.

\end{document}